# A high-speed heterogeneous lithium tantalate silicon photonics platform

Margot Niels[1,2,*,†], Tom Vanackere[1,2,*,†], Ewoud Vissers[1,2], Tingting Zhai[1,2], Patrick Nenezic[1,2], Jakob Declercq[2,3], Cédric Bruynsteen[2,3], Shengpu Niu[2,3], Arno Moerman[2,3], Olivier Caytan[2,3], Nishant Singh[2,3], Sam Lemey[2,3], Xin Yin[2,3], Sofie Janssen[2], Peter Verheyen[2], Neha Singh[2], Dieter Bode[2], Martin Davi[2], Filippo Ferraro[2], Philippe Absil[2], Sadhishkumar Balakrishnan[2], Joris Van Campenhout[2], Günther Roelkens[1,2], Bart Kuyken[1,2,*], Maximilien Billet[1,2,*]

[1] *Department of Information Technology (INTEC), Photonics Research Group, Ghent University–imec, 9052 Ghent, Belgium.*
[2] *imec, Kapeldreef 75, 3001 Leuven, Belgium.*
[3] *Department of Information Technology (INTEC), IDLab, Ghent University–imec, 9052 Ghent, Belgium.*
[†] *These authors contributed equally to this work.*
* *Margot.Niels@UGent.be, Tom.Vanackere@UGent.be, Bart.Kuyken@UGent.be, Maximilien.Billet@UGent.be*

**The rapid expansion of cloud computing and artificial intelligence has driven the demand for faster optical components in data centres to unprecedented levels. A key advancement in this field is the integration of multiple photonic components onto a single chip, enhancing the performance of optical transceivers. Here, silicon photonics, benefiting from mature fabrication processes, has gained prominence in both academic research and industrial applications. The platform combines modulators, switches, photodetectors and low-loss waveguides on a single chip. However, emerging standards like 1600ZR+ potentially exceed the capabilities of silicon-based modulators. To address these limitations, thin-film lithium niobate has been proposed as an alternative to silicon photonics, offering a low voltage-length product and exceptional high-speed modulation properties. More recently, the first demonstrations of thin-film lithium tantalate circuits have emerged, addressing some of the disadvantages of lithium niobate enabling a reduced bias drift and enhanced resistance to optical damage. As such, making it a promising candidate for next-generation photonic platforms. However, a persistent drawback of such platforms is the lithium contamination, which complicates integration with CMOS fabrication processes. Here, we present for the first time the integration of lithium tantalate onto a silicon photonics chip. This integration is achieved without modifying the standard silicon photonics process design kit. Our device achieves low half-wave voltage (3.5 V), low insertion loss (2.9 dB) and high-speed operation (> 70 GHz), paving the way for next-gen applications. By minimising lithium tantalate material use, our approach reduces costs while leveraging existing silicon photonics technology advancements, in particular supporting ultra-fast monolithic germanium photodetectors and established process design kits.**

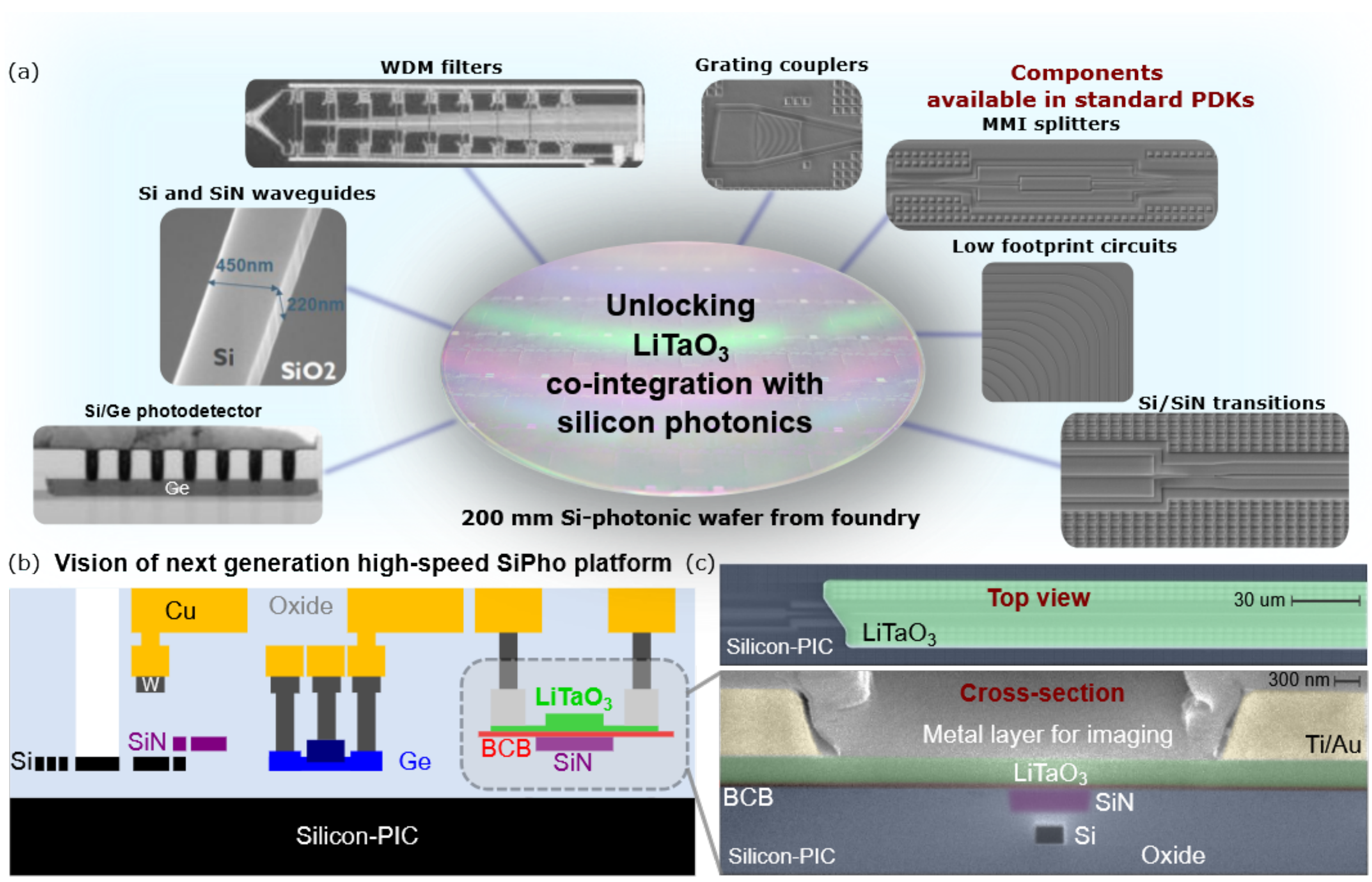

FIG. 1. Next generation high-speed silicon photonic platform. (a) Generic view of a state-of-the-art Si-PIC platform with its basic components. (b) View of the vision of a high-speed Si-PIC's cross-section: next to the components already offered, lithium tantalate modulators need to be integrated. (c) SEM (colourised) picture after the integration of $LiTaO_3$ on a Si-PIC and FIB cross-section (colourised) of a heterogeneous electro-optic device on a Si-PIC.

The exponential growth in data network traffic leads to hardly sustainable power consumption levels in data centres. Addressing this challenge, optical fibres provide an efficient solution for data transport within and between these data centres. Photonic integration, initially developed on indium phosphide[1] and more re-cently on silicon photonic platforms, has significantly enhanced data stream scaling while minimising power consumption[2]. Silicon photonic integrated circuits (Si-PICs) enable the dense integration of complex functionalities, leveraging CMOS-compatible fabrication for high-volume, high-yield, and low-cost production. A typical silicon photonic platform (see Fig. 1) integrates several key components, including Si waveguides with dopants for high-speed modulation, heaters for efficient low-speed thermal tuning, a silicon nitride (SiN) layer for low-loss propagation and high-quality filtering, and germanium photodetectors. The latter have demonstrated bandwidths exceeding 200 GHz[3]. However, emerging standards such as 1600ZR+[4] may require baud rates surpassing 200 GBd, which exceed the performance limits of current Si photonic modulators.

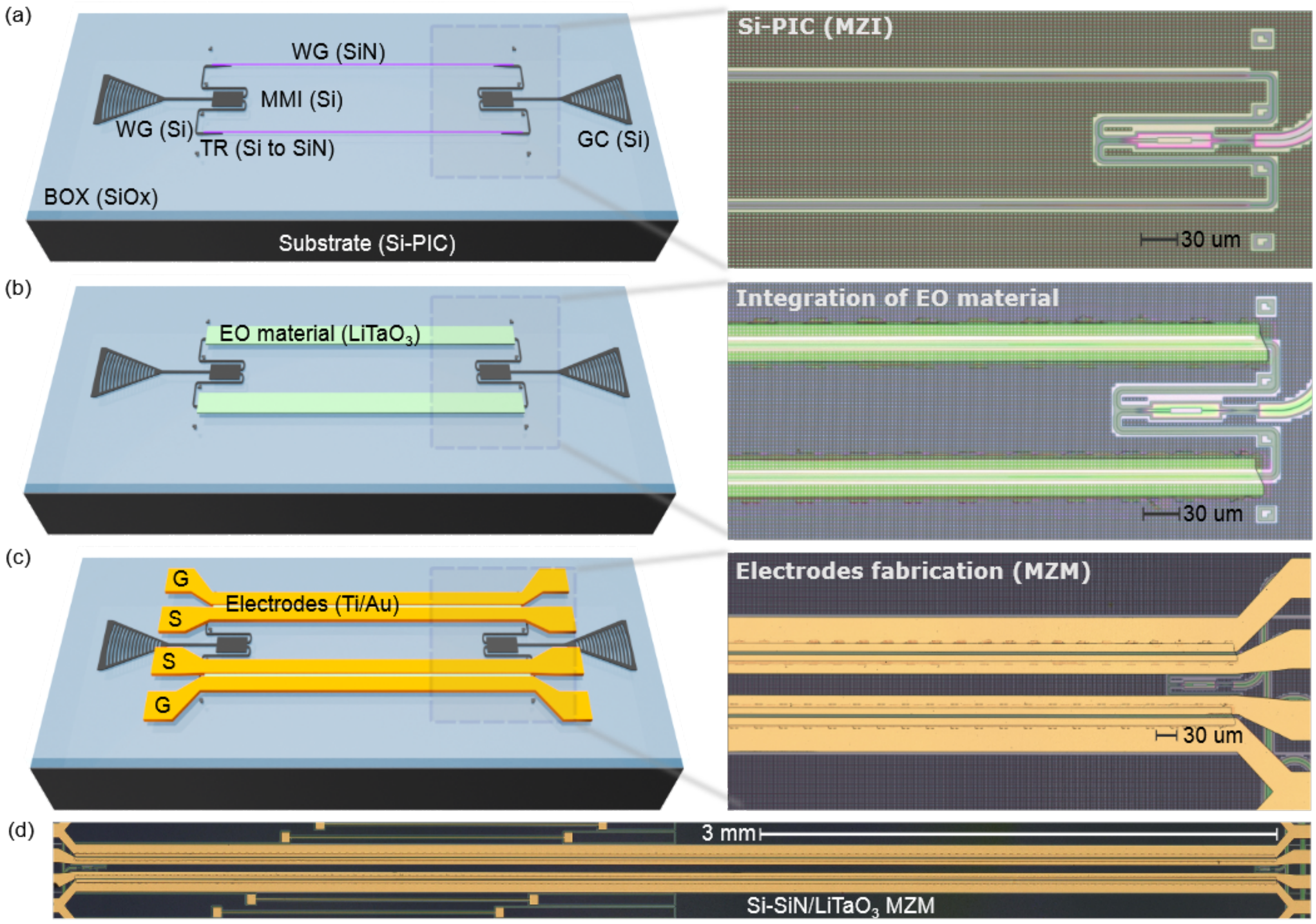

FIG. 2. Integration flow of $LiTaO_3$ on a Si-PIC. (a) Architecture of the passive Si and SiN circuitry and corresponding optical microscope picture. WG: waveguide, BOX: buried oxide, TR: transition, GC: grating coupler, MMI: multimode interferometer, MZI: Mach-Zehnder interferometer. (b) Architecture of the hybrid modulator after the integration of the $LiTaO_3$ and corresponding optical microscope picture. (c) Architecture of the Mach-Zehnder modulator after the fabrication of the electrodes and corresponding optical microscope picture. G: ground, S: signal. (d) Overview of a 7-mm-long heterogeneous Si and SiN/$LiTaO_3$ MZM.

To overcome this limitation, various alternative platforms are under investigation, with the heterogeneous integration of novel electro-optic (EO) materials gaining significant attention. Barium titanate (BTO) presents a promising option due to its capability for direct growth on Si layers and its compatibility with CMOS fabrication processes[5]. While high-speed operation has been achieved, challenges such as high permittivity dispersion and the requirement for a constant bias persist. Alternative approaches include the integration of organic materials[6], plasmonics[7–9], III-V semiconductors[10,11], and graphene[12–16]. However, thin-film lithium niobate (TFLN) has emerged as a standout material, offering a combination of low optical loss and a strong, fast EO coefficient[17]. In demonstrations, bandwidths exceeding 100 GHz[18] have been achieved. Moreover, lithium niobate ($LiNbO_3$) has been heterogeneously integrated onto Si and SiN platforms: techniques such as wafer bonding[19,20] and micro-transfer printing have been successfully employed[21].

Very recently, lithium tantalate ($LiTaO_3$) has emerged as an alternative to $LiNbO_3$[22–25]. The material has a similar EO coefficient as $LiNbO_3$. Yet, it shows a much weaker photorefractive effect and a higher damage threshold. DC stable $LiTaO_3$ EO modulators have been demonstrated[26]. So far, only demonstrations on a monolithic platform have been shown. As with monolithic $LiNbO_3$ optical waveguide platforms, however, the route towards high volume production in CMOS fabs is less clear due to the contamination caused by lithium[27]. Moreover, the integration of other components such as high-speed detectors is not obvious at the moment.

In this work, we present the first demonstration of the heterogeneous integration of $LiTaO_3$ onto a Si-PIC. The integrated modulator is compatible with the Si-PIC's process design kit (PDK), ensuring full compatibility with platform components and enabling their optimal functionality. Leveraging the existing PDK, we achieve seamless integration of EO components without compromising the performance of the established ones. Using a back-end integration approach, based on micro-transfer printing, we ensure compatibility with the entire wafer stack, facilitating co-integration with critical components such as heaters, filters, and germanium photodetectors. A conceptual view is illustrated in Fig. 1. Moreover, the printing technique allows for reversing the poling axis such that complex electrode designs can be realised, allowing for differential driving, such as ground-signal-signal-ground (GSSG), which is hardly doable on monolithic $LiNbO_3$ platforms or wafer bonded devices. Hence, such modulators can be driven by differential drivers with improved linearity thanks to the cancellation of even order harmonics[28]. Moreover, these differential drivers also have improved tolerance to electromagnetic interference (EMI) or power supply noise, especially important in multi-channel devices. The unbalanced Mach-Zehnder modulator (MZM), presented in this work, achieves a measured bandwidth exceeding 70 GHz (limited by measurement equipment) and is estimated to reach 90 GHz. It also exhibits a voltage-length product of 2.3 V·cm, comparable to recent results from monolithic $LiTaO_3$ platforms[24]. Performance can be further enhanced through the optimisation of the hybrid SiN/$LiTaO_3$ cross-section. Furthermore, the hybrid phase modulator in the arms of the Mach-Zhender has a total insertion loss of 2.9 dB, dominated by metal losses caused by a fabrication misalignment (1.6 dB). Finally, data modulation up to 190 GBd is successfully demonstrated highlighting the potential of this approach for next-generation high-speed photonic applications.

## RESULTS

### Design and fabrication of the $LiTaO_3$ MZM

The device architecture relies on an MZM implemented on a cutting-edge Si-PIC platform (imec iSiPP200). A subset of the commercial platform is used for device layout and circuit routing. Here, only the silicon and silicon nitride waveguide layers are processed. Starting from a silicon-on-insulator (SOI) wafer with a 220 nm device layer, the Si layer is patterned with two etch depths (70 nm and 220 nm). Next, the wafer is planarised and a SiN layer with a thickness of 300 nm is deposited. This layer is patterned (with a full etch of 300 nm), and cladded in oxide, after which the wafer is planarised. The routing and splitting of the modulator is done in the silicon waveguide layer. The 50:50 splitter and combiner in the MZM are 1x2 multi-mode interferometers (MMIs), which are available in the PDK. Light is coupled in and out of the chip using grating couplers (GCs), and routing is performed by single-mode waveguides, which are included in the PDK. The schematic of the passive device architecture and an optical microscope picture of the fabricated circuit are depicted in Fig. 2a.

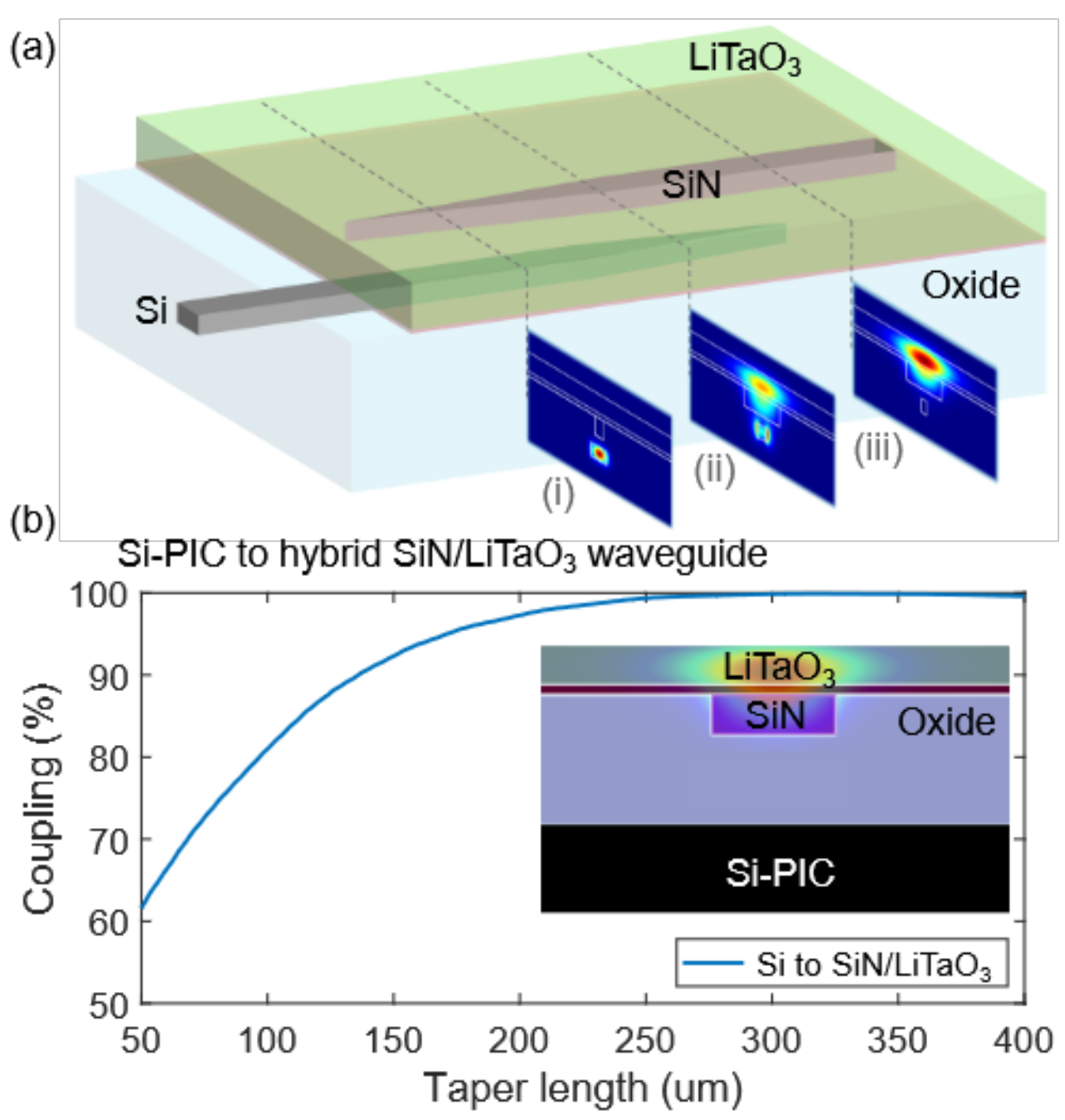

FIG. 3. Transition from the Si waveguide to the hybrid modulator. (a) Schematic of the tri-layer adiabatic transition: mode profiles in the regions of interest are added. (b) Expected transmission from a fully etched Si waveguide to a hybrid SiN/$LiTaO_3$ phase modulator, (inset) and cross-section of the passive stack of an MZM arm.

In a back-end heterogeneous integration step, X-cut thin-film $LiTaO_3$ is incorporated onto both arms of the MZI, hence enabling hybrid phase modulators. The integration of this EO material is achieved using micro-transfer printing technology[29,30].The micro-transfer printing process allows for the transfer of thin films with a large length-to-width ratio. Here, a commercial LTOI wafer is used as the source for the transfer, on which suspended $LiTaO_3$ membranes are prepared. These measure 300 nm in thickness, 30 µm in width, and 7 mm in length. Details on the fabrication process flow of the suspended membranes and the back-end integration are provided in the Methods section.

These membranes are then transferred to the photonic chip using a commercial micro-transfer printer, after applying a 50-nm-thick layer of BCB on the chip as an adhesive. The high membrane density on the source wafer allows for the fabrication of over 12000 membranes (7 mm x 30 µm) on a 4-inch wafer, showcasing the efficient use of the costly electro-optic material. A schematic of the device and the corresponding optical microscope pictures after the successful integration of the $LiTaO_3$ on the platform are illustrated in Fig. 2b. Utilising the versatility of the micro-transfer printing approach, the modulator operates in a push-pull configuration, with the printed $LiTaO_3$ membranes oriented 180° relative to one another. This allows the use of integrated electronic drivers with a differential output[28].

After integration, the final metallisation step involves depositing a 20 nm layer of titanium (Ti) followed by a 1 µm layer of gold (Au) using a lift-off process. This step is required to form the metal electrodes configured in a ground-signal-signal-ground (GSSG) arrangement on top of the device arms. The schematic of the finalised MZM, along with a detailed optical microscope view are shown in Fig. 2c. The entire modulator is depicted in Fig. 2d.

A critical feature of the device is the efficient coupling of light from the Si waveguide to the hybrid EO section of the MZM. This coupling is achieved through an adiabatic transition. In the design, both the Si and the SiN layers of the Si photonic platform are utilised to minimise the coupling loss into the EO structure. A schematic of the coupling structure, along with the optical mode profiles for each region of interest, is shown in Fig. 3. With the vertical tri-layer stack of Si/SiN/$LiTaO_3$, a transition is designed to be robust against fabrication geometry variations, e.g. the layer thickness of SiN and $LiTaO_3$ or waveguide dimensions. The simulated transmission efficiency, depicted in Fig. 3.b, is computed using the eigenmode expansion (EME) method implemented in the commercial software Ansys Lumerical MODE, for a range of taper lengths, and shows that a coupling efficiency of virtually 100 % is possible in this hybrid structure. Further details on the adiabatic transition design are provided in Methods. Excluding the coupling section, the effective length of the active EO SiN/$LiTaO_3$ waveguide is reduced to 6.6 mm, compared to the original 7 mm length of the modulator arms, since a 200 um transition length was chosen.

### Performances of the modulator

The fabricated MZMs are characterised both optically and electro-optically. Initial characterisation involves a transmission wavelength sweep in the O-band, spanning from 1300 nm to 1325 nm. Continuous-wave (CW) optical power is coupled to the chip using a tunable laser. Through an optical fibre and a polarisation controller, the transverse electric (TE) mode is exited in the waveguide via a grating coupler. The optical power transmitted through the MZM is recorded using a power meter, while on the same structure set, a reference waveguide transmission sweep is conducted to account for the contributions of the grating couplers and Si routing.

The resulting normalised MZM transmission spectrum is plotted in Fig. 5.a, where the interference pattern reveals a spectral interferometric pattern with a free spectral range (FSR) of 4 nm and an extinction ratio (ER) of 28 dB at a wavelength of 1310 nm. However, beyond the 3.8 dB insertion loss of the MZM structure defined by the platform PDK, the high-speed electro-optic modulation section introduces an additional 2.9 dB insertion loss, including coupling (0.6 dB) and propagation losses (0.7 dB). The remaining loss is attributed to a misalignment of around 0.6 µm of the final gold layer, which introduces an additional 1.6 dB of optical loss. Further details on the losses of the MZM are available in the Method section.

Next, the device is operated in a quasi-DC modulation experiment. The transmitted CW optical signal is modulated at low electrical frequencies to characterise the half-wave voltage ($V_\pi$). The device is driven by a radio frequency (RF) triangular wave at 100 kHz, generated by an RF signal generator, while the optical wavelength of 1309.26 nm is chosen to correspond to the quadrature operating point of the modulator, enabling operation in the linear modulation regime. The electrical signal is applied to the device using an RF probe and the optical power transmitted through the device is recorded using a photodetector and an oscilloscope. This result is then plotted as a function of the applied voltage. The results, shown in Fig. 5.b, reveal a $V_\pi$ of 7.0 V for a single phase shifter (one arm of the MZM), which corresponds to 3.5 V in the push-pull amplitude modulation configuration (both arms of the MZM). This yields a voltage-length product ($V_\pi L$) of 2.3 V·cm, consistent with recent reports on $LiTaO_3$ MZMs fabricated on the $LiTaO_3$-on-insulator (LTOI) platform[24,26]. Moreover, when comparing this result to the expected $V_\pi$ curve, calculated using finite element method (FEM) simulations in combination with an analytical model, it demonstrates excellent agreement with the measurement as depicted in the overlay in Fig. 5.b. More information on the simulation of $V_\pi$ is available in Methods.

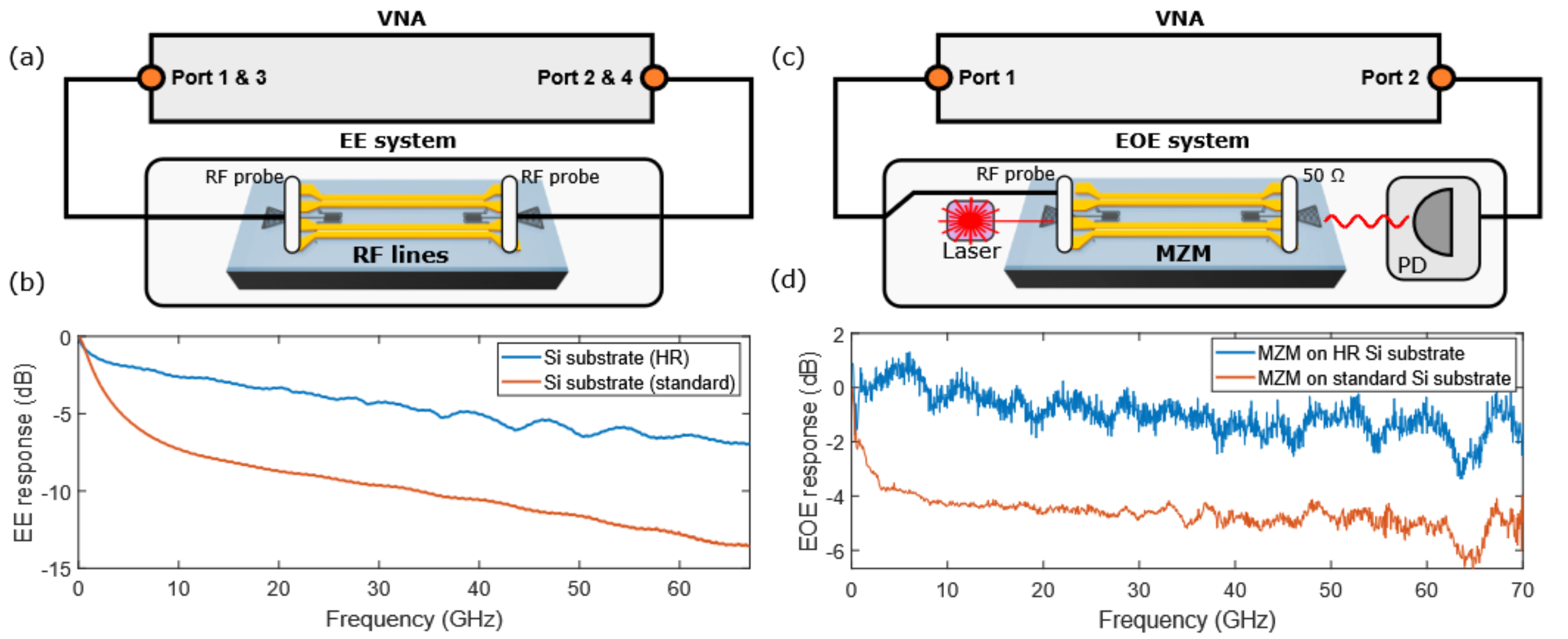

FIG. 4. High-speed characterisation of the modulator. (a) Measurement setup (EE system) for the RF line frequency response characterisation. (b) Results showing the comparison using regular Si substrate and high resistivity substrate. (c) Measurement setup (EOE system) for the modulator frequency response characterisation. (d) Results showing the comparison for a modulator integrated on a platform using a regular Si substrate and high resistivity substrate.

The high-speed response of the device is characterised in twofold: first, the electrodes are characterised by performing an electrical-to-electrical (EE) measurement and afterwards, the full electrical-to-optical-to-electrical (EOE) response is measured. The EE responses are measured with a vector network analyser (VNA) connected to the chip with a set of GSSG probes, as shown in Fig. 4.a. This measurement is carried out on two different chips. On the first chip, the RF lines are placed to drive an MZM which is fabricated on a wafer with a regular Si substrate. In the second case, a high-resistivity Si substrate is used for the wafer. The comparison between the two lines, presented in Fig. 4.b, demonstrates a clear enhancement of the RF bandwidth when using the high-resistivity Si substrate due to lower absorption loss of the RF wave in the substrate.

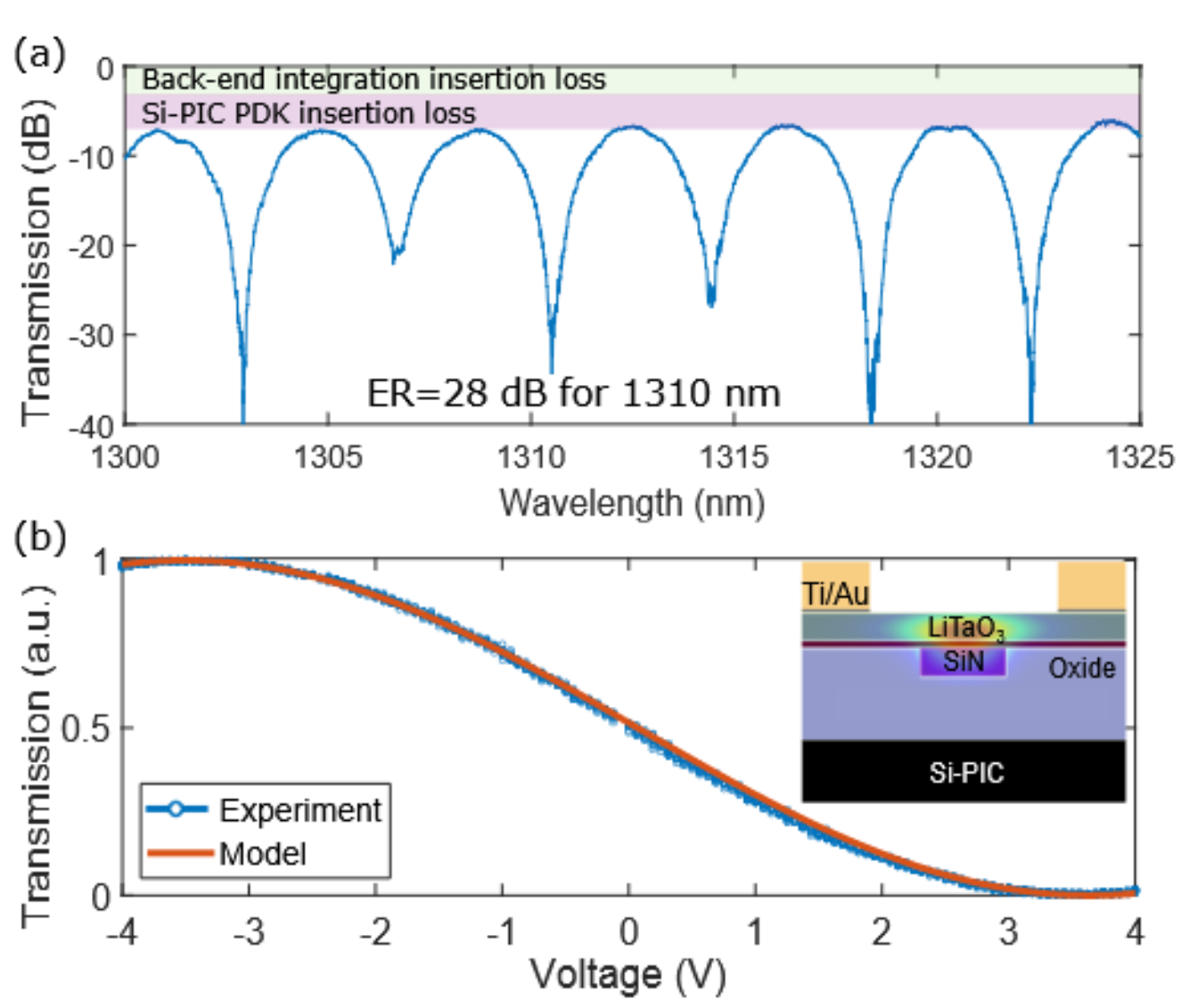

FIG. 5. Quasi-DC characterisation of the modulator. (a) Measurement of the modulator transmission as a function of wavelength (b) Normalised transmission of the modulator as a function of applied voltage as well as the simulated response, and (inset) cross-section of the full stack of an MZM arm.

The EOE response of the MZMs, fabricated on a standard and a high resistivity Si substrate, are then measured using a similar experimental setup (Fig. 4.c). The device is connected to one port of the VNA using a GS probe on one side, and the opposite side of the device is terminated with a 50 Ω load, to prevent electrical reflections at the far side. CW optical power is coupled from the laser to the Si-PIC with an optical fibre, transmitted through the MZM, out-coupled and converted into an electrical signal by a high-speed photodetector. The resulting signal is received at the second port of the VNA. The EOE frequency response of the MZMs are presented in Fig. 4.d for both modulator types. Again, the device with a high resistivity substrate demonstrates a much higher 3-dB cut-off frequency in excess of 70 GHz, limited by the experimental setup (the bandwidth of the VNA and the photodetector).

## Data transmission

As a demonstration, the device is utilised in a link to transmit data for which the setup is presented in Fig. 6.a. A differential signal, generated by a 256 GS/s arbitrary waveform generator (AWG), is applied to the device using a GSSG probe. A second probe is employed to terminate the device on the opposite side of the transmission line with two 50 Ω loads to prevent electrical reflections. The optical signal is amplified using a praseodymium-doped fibre amplifier (PDFA) before being detected by a high-speed photodetector. The detected signal is subsequently recorded using a sampling oscilloscope. To optimise the transmission, the link is characterised to facilitate the equalisation of the signal generated by the AWG. As such, clear eye diagrams are generated. Supplementary Data Fig. 3 shows the measured frequency response of the data transmission link (magnitude in a. and phase in b.), which is compensated at the AWG. A noticeable performance drop-off is observed in the link characterisation around 70 GHz, attributed to the bandwidth limitations of the electrical cables, RF probes, the high-speed photodetector and the AWG. All transmitted signals are based on standardised pseudorandom bit sequences (PRBS-15).

A 64-tab feed-forward equalisation (FFE) algorithm is applied at the receiver end, using the sampling oscilloscope's software to enhance the received signal. Fig. 6.b presents the measured bit error rate (BER) derived from the generated eye diagrams using both non-return-to-zero (NRZ) and pulse-amplitude modulation 4-level (PAM4) formats. A dashed line indicates the BER threshold for a forward error correction (FEC) algorithm with an overhead of 15 %. The lower limit of the plot, at $3\text{x}10^{-4}$, indicates the error rate threshold below which statistically significant assessments could not be made, due to the limited number of bits transmitted during the demonstration. Fig. 6 also includes representative eye diagrams, all of which incorporate FFE.

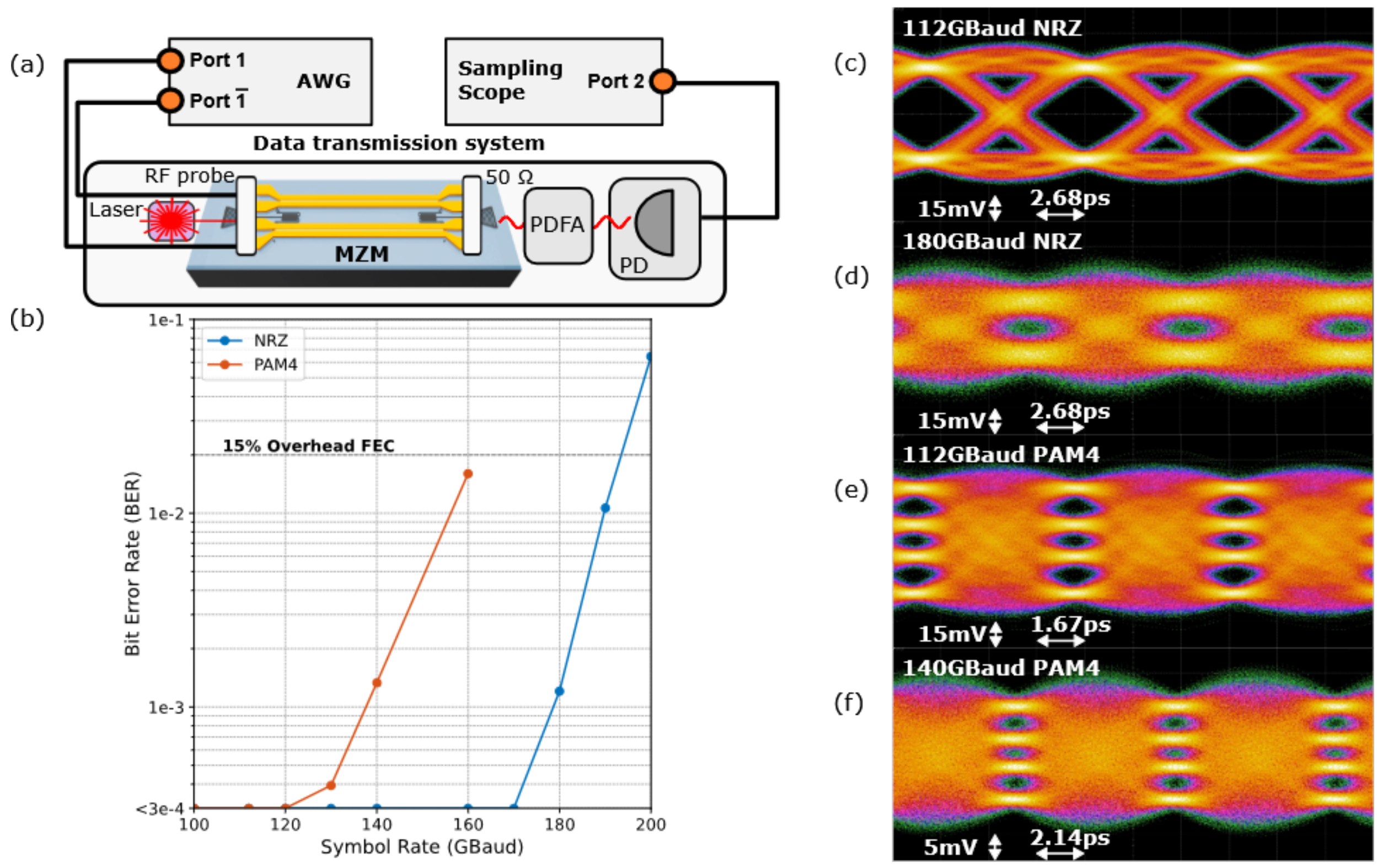

FIG. 6. Data transmission experiment. (a) Measurement setup for eye-diagram measurements. (b) Bit error rate (BER) in the transmitted bit sequence for NRZ (blue) and PAM4 (red) measured signals. A horizontal dashed line indicates the threshold for forward error correction (FEC) with an overhead of 15 %. Eye diagrams for (c) 112 GBaud NRZ, (d) 180 GBaud NRZ, (e) 112 GBaud PAM4, (f) 140 GBaud PAM4, all figures used correction at the transmitter side and a 64 tap FFE on the receiver side, however, no offline digital post-processing was done

## DISCUSSION

### Performance improvement

High-speed hybrid EO/Si-PIC modulators are designed for integration into data communication systems, encompassing both inter- and intra-data centre links to enable massive data transmission. In such systems, power consumption is a critical parameter. The presented devices demonstrate significant potential for further optimisation in this regard, particularly through the reduction of electrical driving power.

Recently, high-quality etching of $LiTaO_3$ has been demonstrated[24]. By employing this etching process, the $LiTaO_3$ layer in the hybrid modulator can be patterned. The patterned $LiTaO_3$ layers would allow for better confinement and hence a stronger EO effect as the electrode separation can be reduced (see Methods). Consequently, the gap can be reduced from 5.5 µm to 3.4 µm, resulting in a stronger RF field and a lower simulated $V\pi$ of 2.7 V, corresponding to a simulated voltage-length product of 1.8 V·cm. As discussed in the Method section, for this configuration efficient adiabatic couplers can be made on the platform that allows for high coupling efficiency of more than 99 %. Moreover, proper design allows similar very high coupling efficiencies (> 99 %) for transverse misalignments of the $LiTaO_3$ ridge waveguide of up to 0.5 µm. These misalignments remain within the alignment accuracy specifications of wafer-scale micro-transfer printing tools.

## CONCLUSIONS

The successful heterogeneous integration of lithium tantalate onto a silicon photonic platform with minimal insertion loss is demonstrated. Implemented on a standard 200-mm Si-PIC platform, our approach ensures seamless integration with existing platform PDK components while preserving their performance. By leveraging back-end integration, we maintain compatibility across the entire wafer stack, enabling efficient co-integration with critical components such as waveguides, heaters, filters, and germanium photodetectors. The manufactured MZM achieves a measured bandwidth exceeding 70 GHz, limited by the measurement setup, and exhibits a voltage-length product of 2.3 V·cm comparable with demonstrated LTOI platforms. The hybrid phase modulators in the MZM arms induce an additional 2.9 dB insertion loss. Lastly, high-speed transmission experiments are performed, reaching symbol rates of 190 Gbaud (NRZ) and data rates of more than 320 Gbit/s (PAM4).

These results mark a significant step toward high-speed, high-performance hybrid lithium tantalate photonic platforms. They pave the way for scalable, industry-compatible electro-optic integration for low-energy consumption data communication links that can be fabricated in high volumes.

## Data Availability

The datasets generated during and/or analysed during the current study are available from the corresponding author on reasonable request.

## Acknowledgments

The authors would like to acknowledge the contribution of imec's 200 mm pilot line for silicon photonics wafer fabrication and imec's PDK team for the mask tape-out. The authors also would like to acknowledge the contribution from Priya Eswaran and Semih Culhaoglu for the help in Si-PIC process development, Steven Verstuyft, Peter Geerinck, Elif Özçeri, Liesbet Van Landschoot for the help during the lithium tantalate device fabrication, and Clemens Krückel and Joris Van Kerrebrouck for measurement support.

## Author contributions

S.B., P.A., J.C., S.J., P.V., N.S.[2], D.B., M.D. and F.F. developed and fabricated the silicon nitride, silicon platform on 200 mm. M.N., T.Z. and M.B. fabricated the hybrid $LiTaO_3$/SiN devices and realised the heterogeneous integration. E.V., A.M., P.N. and O.C.

performed numerical simulations. E.V., A.M. and T.V. designed the devices. M.N. and T.V. performed the characterisation (optical and quasi-dc). M.N., T.V., N.S[2,3], J.D., C.B. and S.N. performed the high-speed characterisation and data communication experiment. M.N., T.V., B.K. and M.B. prepared the figures, data, and the manuscript with input from other authors. P.A., S.L., X.Y., G.R., B.K. and M.B. supervised the project:

## Competing interests

The authors declare no conflict of interest.

## Funding

We want to thank the European Space Agency for funding under the E/0365-70 - NAVISP, LEO Project and The Research Foundation Flanders (FWO) for project 3G035722 and the FWO and F.R.S.-FNRS under the Excellence of Science (EOS) program (40007560).

## SUPPLEMENTARY MATERIAL

### Process flow for heterogeneous $LiTaO_3$ integration

The fabrication of suspended $LiTaO_3$ membranes starts with a 300 nm $LiTaO_3$, on 2 µm oxide, on a Si substrate wafer (Supplementary Data Fig. 1.a). The $LiTaO_3$ is patterned into rectangles of 30 µm x 7 mm with UV-lithography, using an amorphous silicon (aSi) hard mask suitable for an argon (Ar) based reactive ion etching (RIE). The aSi hard mask is then removed in a potassium hydroxide (KOH) solution (Supplementary Data Fig. 1.b). As a second step, the oxide layer is patterned with UV-lithography and RIE, after which the oxide is etched down to the Si substrate (Supplementary Data Fig. 1.c). A mechanical encapsulation layer is then added, consisting of a photoresist layer exposed with UV-lithography which is subsequently developed. The patterns in the photoresist encapsulation are made in order to define mechanical links from the Si substrate to the $LiTaO_3$ membranes, allowing for the suspension of $LiTaO_3$ when the devices are released (Supplementary Data Fig. 1.d). The last step is the releasing of the $LiTaO_3$ membranes in a wet etchant: buffered hydrofluoric (BHF) acid (Supplementary Data Fig. 1.e). The picture from Supplementary Data Fig. 1.f shows an overview of the final sample. In this picture, 21 membranes of 30 µm x 7 mm are located on a sample of 2 mm x 7 mm. The membrane density is 1.5 membranes/mm$^2$, and for a 4-inch wafer, the total number of membranes would be more than 12 000.

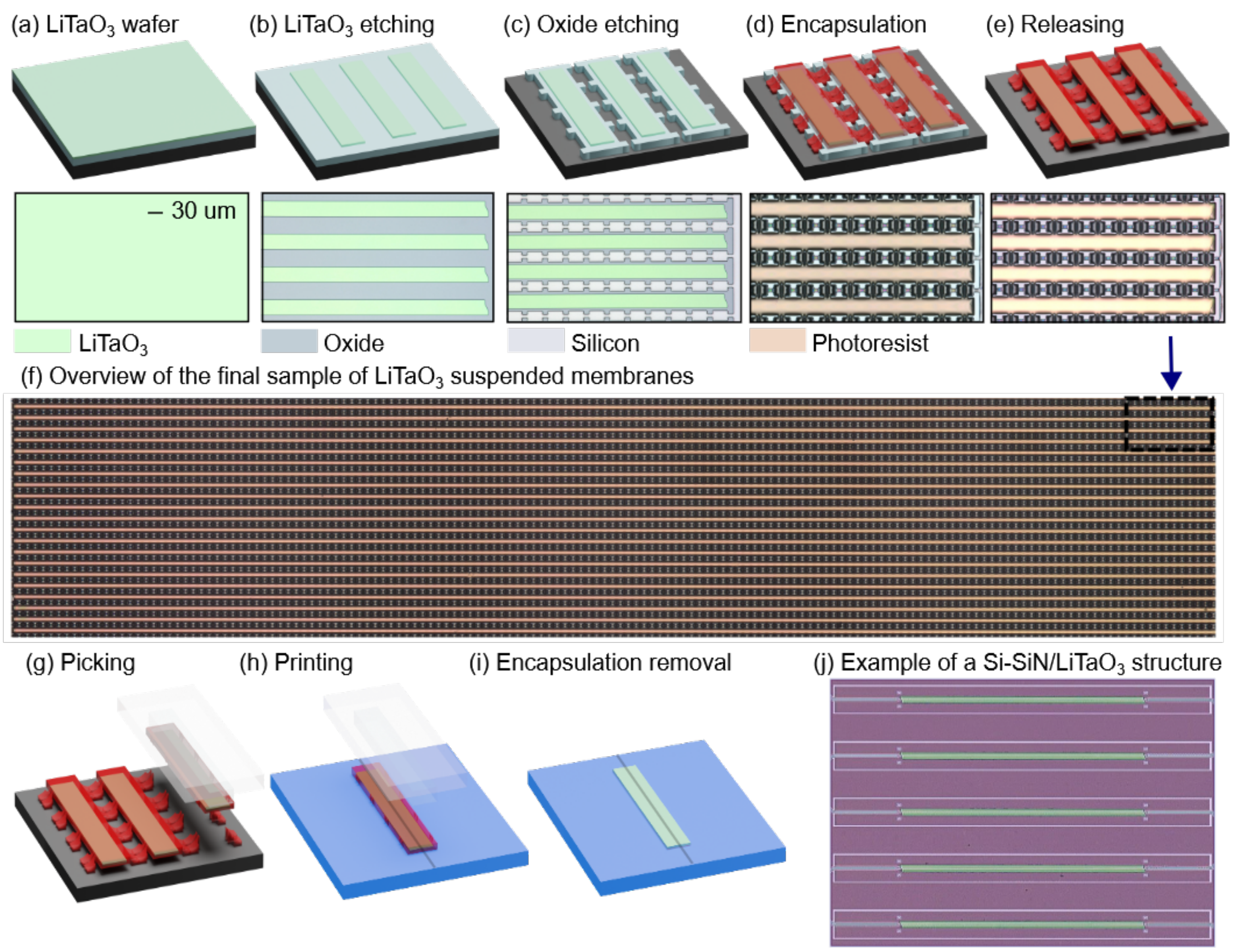

Supplementary Data Fig. 1. Description of the successive steps for the fabrication and printing of suspended $LiTaO_3$ membranes. (a) $LiTaO_3$ (300 nm) / oxide (2 µm) / Si wafer (substrate) starting point. (b) Patterning of the $LiTaO_3$ membranes. (c) Patterning of the oxide release layer. (d) Photoresist mechanical encapsulation of the structures. (e) Undercutting (releasing) of the oxide layer, making the structures suspended. (f) Overview of the final sample with suspended $LiTaO_3$. (g) picking of the suspended $LiTaO_3$. (h) Printing of the $LiTaO_3$ on a pre-processed external chip. (i) Encapsulation removal. (j) Overview of a few successfully printed $LiTaO_3$ membranes.

At this step, the membranes are ready to be integrated on a different substrate. They are picked using a polymer PDMS stamp driven by a commercial micro-transfer printer tool (Supplementary Data Fig. 1.g). For better adhesion, a BCB buffer layer is applied to the target before printing. The membranes are then placed on the host substrate, and the stamp is retracted, leaving the membrane at its final location (Supplementary Data Fig. 1.h). The photoresist encapsulation is subsequently removed using an RIE oxygen plasma (Supplementary Data Fig. 1.i), after which the BCB layer is cured at 280°C for 90 min. An example of a Si chip with a cascade of 1-mm-long $LiTaO_3$ sections used as an insertion loss test structure is depicted in Supplementary Data Fig. 1.j.

On top of the loss characterisation test structures, the integration of $LiTaO_3$ membranes is done for four MZM devices. To illustrate the stability of the process flow, the transmission as a function of the wavelength of the devices is recorded and presented in Supplementary Data Fig. 2. The four devices, combining eight $LiTaO_3$ integrated membranes in total, show similar results with an ER $> 25$ dB for the peak near 1310 nm. The wavelength sweeps in Supplementary Data Fig. 2 are normalised by subtracting the grating couplers and routing outside the actual MZM.

### Loss characterisation

Two kinds of optical loss contributions can be distinguished: propagation loss through the hybrid SiN/$LiTaO_3$ waveguide and the coupling or transition loss to get light from a Si waveguide to a hybrid SiN/$LiTaO_3$ waveguide (see adiabatic coupling for more details on this transition). To determine the propagation loss, hybrid SiN/$LiTaO_3$ waveguides with different lengths are fabricated. These are indicated as S1 and S2 in Supplementary Data Table. I and have a propagation length of respectively 0.66 and 6.66 mm. For the transitions, two structures are prepared where

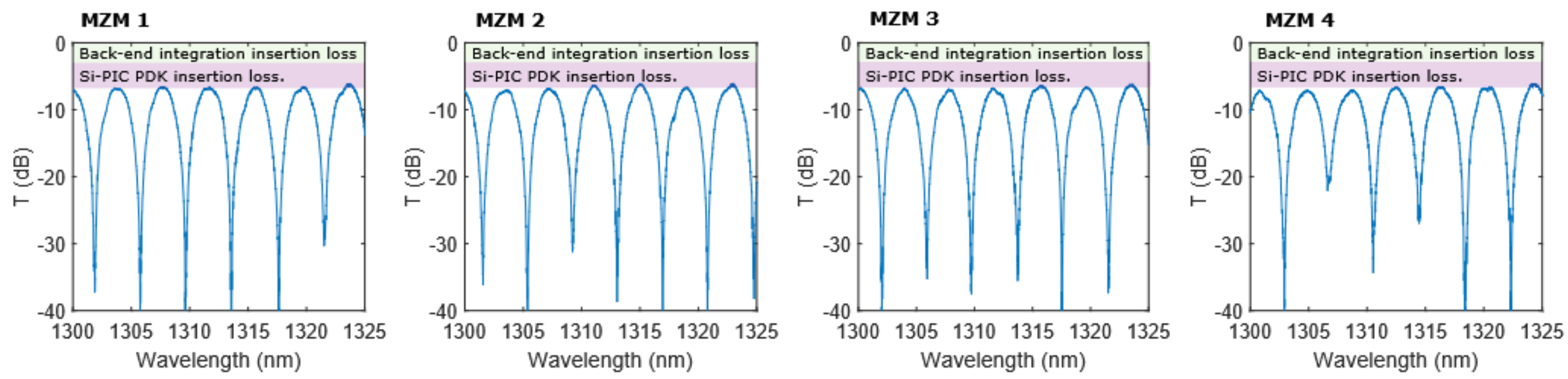

Supplementary Data Fig. 2. Transmission (T) measurements for four fabricated MZMs. The losses from the grating couplers and external routing are removed. The insertion loss contribution from the Si-PIC components (C1 and C2 from Supplementary Data Table. II) and the contribution of the back-end integration process (C3, C4 and C5 from Supplementary Data Table. II) are described on top of the graphs.

one has only two transitions (one in, one out) and the other one has 28 transitions. Those are indicated as S3 and S4 in Supplementary Data Table. I and have, in total, the same propagation length. Comparing the transmission of all structures, and taking the propagation loss in the Si waveguides into account, a propagation loss of 1.0 ± 0.5 dB/cm and a transition loss of 0.3 ± 0.1 dB per transition can be deduced.

| Test structure | Amount of membranes (#) | Length of propagation in SiN/$LiTaO_3$ (mm) | Amount of transitions (#) | Si propagation loss (dB) | Transmission at 1310 nm (dB) |
|---|---|---|---|---|---|
| S1 | 1 | 0.6 | 2 | 0.2 ± 0.01 | -8.2 ± 0.2 |
| S2 | 1 | 6.6 | 2 | 0.2 ± 0.01 | -8.8 ± 0.2 |
| S3 | 1 | 0.3 | 2 | 0.4 ± 0.01 | -8.1 ± 0.2 |
| S4 | 1 | 0.3 | 28 | 2.5 ± 0.04 | -16.8 ± 0.2 |

Supplementary Data Table. I. Description and measured transmission for all test structures needed to determine the propagation loss and transition loss of the $LiTaO_3$ slabs.

| Label | Component | Loss in the MZM |
|---|---|---|
| C1 | MMI splitters | 0.8 ± 0.2 |
| C2 | Si routing inside MZM | 3.0 ± 0.4 |
| C3 | Transitions (Si to SiN/$LiTaO_3$) | 0.6 ± 0.2 |
| C4 | EO section (SiN/$LiTaO_3$) | 0.7 ± 0.3 |
| C5 | Metal misalignment | 1.6 ± 0.1 |

Supplementary Data Table. II. Measurement of the loss for each component of the MZM allowing to extract the contribution to the insertion loss from the Si-PIC PDK (C1 and C2) and the back-end integration process (C3, C4 and C5).

All component losses inside the MZM are also listed in Supplementary Data Table. II. The losses of the MMI splitter (C1) and Si routing (C2) are measured from the process control structures on the same chip. The transition from Si to SiN/$LiTaO_3$ (C3) and the propagation through the hybrid SiN/$LiTaO_3$ (C4) are extracted from the loss test structures as described above. Also, the introduction of metal traces on top of the $LiTaO_3$ (C5) introduces extra loss. A misalignment of about 600 nm is observed and can be seen in the FIB cut. As no test structures are available to quantify this loss, a finite element method simulation was carried out in COMSOL to identify its contribution to be 1.6 dB.

### Adiabatic coupling from Si platform to the hybrid $LiTaO_3$ modulator

An adiabatic tapering structure is optimised to efficiently couple the light from the Si waveguide to the EO section. The following paragraph clarifies the design of the taper section referred to when describing the fabricated chips (hybrid SiN/$LiTaO_3$-on-Si modulator).

At the start of the transition, light remains confined within the Si waveguide despite the introduction of the $LiTaO_3$ layer, due to the significant refractive index contrast between the two materials as shown in Fig. 3.a. At this stage, the relevant material stack consists of a 380 nm x 220 nm Si waveguide, encapsulated by a 500 nm thick oxide layer and on top of that a 30 µm x 300 nm $LiTaO_3$ layer. Further in the transition structure, a 300 nm SiN layer is introduced, positioned 150 nm above the Si and 70 nm below the $LiTaO_3$ (see (Fig. 3.a.(i)). At this point, the SiN waveguide is tapered from a narrow tip (150 nm), ensuring that the optical mode remains predominantly in the Si, with minimal influence from that SiN taper tip. Subsequently, the SiN waveguide undergoes an adiabatic width expansion, after which the Si waveguide tapers down to 150 nm, facilitating a gradual transition of the optical mode from Si to the hybrid SiN/$LiTaO_3$ configuration (Fig. 3.a.(ii)). The SiN width is then adjusted to its final value of 900 nm to achieve the optimal confinement of the hybrid mode (Fig. 3.a.(iii)), determined as a trade-off between minimising $V_\pi$ by having as much as possible of the light in the $LiTaO_3$ layer and mitigating subsequent metal-induced losses by achieving sufficient confinement. In this configuration, the lateral alignment of the $LiTaO_3$ with the Si/SiN waveguides does not significantly affect the efficiency of the transition. The taper dimensions are optimised using an EME solver in Lumerical, resulting in a final taper length of 200 µm.

### Half wave voltage calculation

The $V_\pi$ simulations are done using a combination of multiple FEM simulations in COMSOL and an analytical model. The EO phase shift ($\Delta\phi$) experienced by an MZM in a push-pull configuration is governed by the following expression:

$$\Delta\phi = \frac{2\pi n_e^4 \Gamma r_{33} V}{n_{eff} \lambda G} L \tag{1}$$

where $n_e = 2.1269$ is the extraordinary refractive index, $\Gamma$ represents the overlap factor between the RF and optical modes, $r_{33} = 30.5$ pm/V is the EO coefficient of $LiTaO_3$, $V$ is the applied voltage, $L = 6.6$ mm is the effective length of one modulator arm, $\lambda = 1310$ nm is the optical wavelength, $n_{eff}$ is the effective refractive index of the optical mode, and $G$ denotes the electrode gap. The simulated parameters ($\Gamma$, $n_{eff}$ and $G$) are listed in Supplementary Data Table. III.

| Parameter | Hybrid EO waveguide | Unit |
|---|---|---|
| $\Gamma$ | 0.454 | - |
| $n_{\text{eff}}$ | 1.852 | - |
| $G$ | 5.5 | $\mu$m |

Supplementary Data Table. III. Parameters for a hybrid EO MZM.

Using the equation and the simulation results, the theoretical values of $V_\pi$ for an EO-section with 6.6 mm length are calculated to be 3.6 V for the slab push-pull configuration, which shows excellent agreement with measurements.

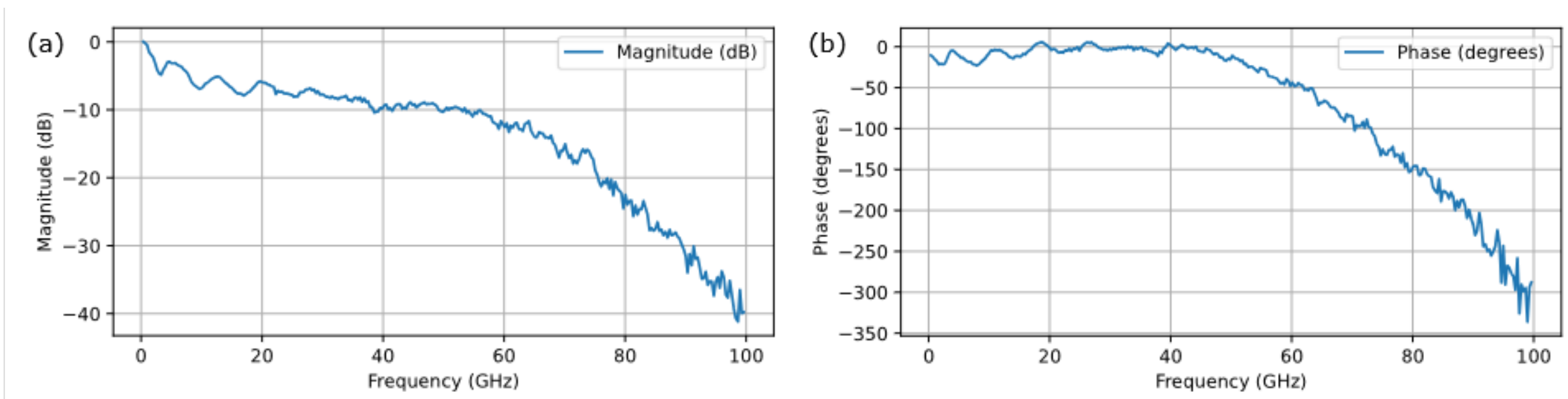

Supplementary Data Fig. 3. Data transmission link characterisation. (a) Amplitude and (b) phase characteristic of the data transmission link, used to generate eye diagrams.